\documentclass[
  reprint,nofootinbib,notitlepage
  amsmath,amssymb,
  aps,pre,raggedbottom
]{revtex4-1}
\usepackage{amsmath}
\usepackage{amssymb}
\usepackage{amsfonts}
\usepackage{amsthm}
\usepackage{bm}
\usepackage{graphicx}

\newcommand{\avg}[1]{\langle#1\rangle}

\newcommand{\oper}[1]{\mathcal{#1}}

\DeclareMathOperator{\mi}{\mathrm{i}}
\DeclareMathOperator{\me}{\mathrm{e}}

\begin{document}

\title{
  Volterra-series approach to stochastic nonlinear dynamics: linear response of the
  Van der Pol oscillator driven by white noise
}
\date{\today / Revision 6-AJH}
\author{Roman Belousov}\email{belousov.roman@gmail.com}
\affiliation{
  Abdus Salam International Centre for Theoretical Physics\\
  Strada Costiera 11, 34151, Trieste, Italy
}
\author{Florian Berger}
\author{A.J. Hudspeth}
\affiliation{
  Howard Hughes Medical Institute, Laboratory of Sensory Neuroscience,
  The Rockefeller University, New York, NY 10065, USA
}
\begin{abstract}
  The Van der Pol equation is a paradigmatic model of relaxation
  oscillations. This remarkable nonlinear phenomenon of self-sustained
  oscillatory motion underlies important rhythmic processes in nature
  and electrical engineering. Relaxation oscillations in a real system
  are usually coupled to environmental noise, which further enriches
  their dynamics, but makes theoretical analysis of such systems and determination
  of the equation's parameter values a difficult task. In a companion paper we have
  proposed an analytic approach to a similar problem for another classical
  nonlinear model, the bistable Duffing oscillator.  Here we extend
  our techniques to the case of the Van der Pol equation driven by
  white noise. We analyze the statistics of solutions and propose a
  method to estimate parameter values from the oscillator's time
  series. We use experimental data of active oscillations in a biological system
  to demonstrate how our method applies to real observations and how it can be generalized
  for more complex models.
\end{abstract}
\maketitle

\section{Introduction}
Balthazar van der Pol introduced the concept of relaxation oscillations together
with his eponymous equation for a simplified dynamics of a triode electric circuit
\cite{vdp1926,vdphistory}. Regarded as a power-series approximation for a more general
class of Lienard systems \cite[Sec. 7.4 and 7.5]{StrogatzII}, this model became a
paradigm of self-sustained oscillatory motion \cite{vdphistory}. Besides its applications
in engineering, the Van der Pol equation and its generalizations are used to describe
various rhythmic processes in biology
\cite{Beta2017,Oates2012,Buzsaki2004,Nomura1993,Rompala2007,Duifhuis1986,Talmadge1990,VanDijk1990,vdp1928,vdp1940,Mirollo1990,Cherevko2016,Cherevko2017,fitzhugh1961,nagumo1962,FitzHughNagumo,Alonso2014,Mindlin2017}.

Self-sustained oscillations are ubiquitous in living systems on different length
and time scales. Examples include intracellular oscillations of molecular concentrations
\cite{Beta2017}, pattern formation and dynamics of tissues \cite{Oates2012}, neuronal
activity \cite{Buzsaki2004,Nomura1993}, circadian clocks \cite{Rompala2007}, otoacoustic
emissions from the ear \cite{Duifhuis1986,Talmadge1990,VanDijk1990}, the beating
of a heart \cite{vdp1928,vdp1940}, the synchronized flashing of fireflies \cite{Mirollo1990},
and hemodynamics \cite{Cherevko2016,Cherevko2017}. Much theoretical work has been
devoted to developing mathematical description for such systems \cite{Roenneberg2008}.
Often the existing models rely on parameters that are difficult to determine from
experimental data. The mathematical description is then limited to qualitative or
conceptual studies.

To facilitate quantitative research we develop analytical techniques for self-sustained
oscillations of the Van der Pol type with a moderate level of noise. In particular,
we derive approximate expressions for the linear response of the Van der Pol oscillator.
In our approach the nonlinear problem of self-sustained oscillations can thus be
mapped onto an effective linear model that reproduces the main features of the original system.

Furthermore, we propose a method to estimate parameter values of the Van der Pol equation
directly from time series, as typically observed in experiments. By fitting empirical
observations to the analytical expressions that we derive, it is straightforward
to determine the underlying model. We demonstrate and validate this approach with
stochastic simulations (Sec.~\ref{sec:num}) and experimental data of active oscillations
from a bullfrog's mechanoreceptive cells in the inner ear (Sec.~\ref{sec:exp}).

A general form of the Van der Pol equation that we consider in this paper extends
the harmonic oscillator by introducing a nonlinear dissipative term in the equation
\begin{equation}\label{eq:main}
  \ddot{x} + a \dot{x} + b x + c \dot{x} x^2 = f
\end{equation}
for an unknown function of time $x(t)$ and an external force $f(t)$; the constants
$a$, $b$, and $c$ are, respectively, the friction coefficient, the stiffness, and
the Van der Pol damping parameter. Because the above equation is of second order in
time, the phase of this system is specified by two degrees of freedom $(x,\dot{x})$.

The self-sustained oscillations of the Van der Pol equation correspond to its limit-cycle
solution. In the absence of the external force $f(t)$, all trajectories of this system
relax to a periodic orbit in the phase space. Self-sustained oscillations exist if
the friction constant in Eq.~(\ref{eq:main}) is negative ($a<0$). The Van der Pol
system is stable when the parameters $b$ and $c$ are both positive. The amplitude
of the limit cycle, which encircles an unstable equilibrium point in the phase space,
shrinks to zero when $a=0$ and disappears for $a>0$. Therefore the Van der Pol oscillator
with $a\ge0$ behaves as a monostable system. This dynamical regime is not studied
in the present paper and should be treated by a different approach \cite[Appendix
A]{Belousov2019I}.

Environmental noise, which intertwines with relaxation oscillations of real systems,
is often modeled by a stochastic force $f(t) = A \dot{w}(t)$, with a constant amplitude
$A > 0$ and Gaussian white noise $\dot{w}(t)$ of zero mean and unit intensity.
One must usually resort to complex measures to determine the model's parameter values
for this class of stochastic nonlinear problems \cite{Alonso2014,Mindlin2017,Cherevko2016,Cherevko2017}.

Previously we demonstrated that time series of a
second-order dynamical system---the stochastic Duffing oscillator---contains enough
information to infer the parameter values of the underlying nonlinear model \cite{Belousov2019I}. Here
we extend our analysis to the case of Van der Pol relaxation oscillations driven
by white noise. After deriving analytical expressions for approximate solutions and time-series
statistics of Eq.~(\ref{eq:main}), we use these formulas to devise a parametric
method of inference.

Our approach is based on the functional series of Volterra \cite{Schetzen,Rugh},
which we expand up to the linear-response term. The analytical results and the inference
method that we propose are therefore applicable to relatively small noise amplitudes
$A$; more details on the system's physical scales are given in Sec.~\ref{sec:num}.
Even in the absence of external driving, the statistical properties of the relaxation
oscillations are far from trivial. This feature of the Van der Pol equation renders
the time-series analysis more difficult than in the case of the Duffing oscillator
\cite{Belousov2019I}. Because we must also approximate the limit-cycle solution of
Eq.~(\ref{eq:main}), for which no closed-form expression is known, our development
is restricted to moderate regimes of driving noise and nonlinear behavior.

\section{\label{sec:thr}Theory}
\subsection{\label{sub:lrsp} Linear response of the Van der Pol oscillator}
A Volterra series is a polynomial functional expansion of the form
\begin{multline}\label{eq:Volterra}
  x(t|f) = x_0(t) + \int_0^t dt_1\, g_1(t-t_1) f(t_1)\\
    + \iint_0^t dt_1 dt_2\, g_2(t-t_1,t-t_2) f(t_1) f(t_2) + ...\\
\end{multline}
in which $g_1(t)$ and $g_2(t)$ are the Volterra kernels of the linear and quadratic
terms in the argument function $f(t)$. Provided that the above series exists, a truncated
expansion~(\ref{eq:Volterra}) approximates solutions of Eq.~(\ref{eq:main}) driven
by a small external force:
\begin{multline}\label{eq:xone}
  x(t) \simeq x_0(t) + \int_0^t dt_1\, g_1(t-t_1) f(t_1)
    \\ = x_0(t) + \gamma_1(t),
\end{multline}
in which we neglect terms of the second and higher orders in $f(t)$. The functions
$x_0(t)$ and $\gamma_1(t)$ can be found by using the variational approach \cite[Sec. 3.4]{Belousov2019I,Rugh},
which yields a set of equations
\begin{eqnarray}
  \label{eq:zero}
	\ddot{x}_0 + a \dot{x}_0 + b x_0 + c\dot{x}_0 x_0^2 &=& 0,
  \\\label{eq:one}
	\ddot\gamma_1 + (a + c x_0^2)\dot\gamma_1 + (b + 2 c \dot{x}_0 x_0)\gamma_1 &=& f,
  \\&\cdots&\nonumber
\end{eqnarray}
Equation~(\ref{eq:zero}), which uniquely defines $x_0(t)$ for a given initial condition
$\big{(}x(0),\dot{x}(0)\big{)}$, is equivalent to the autonomous Van der Pol problem---Eq.~(\ref{eq:main})
with $f\equiv0$. The linear Eq.~(\ref{eq:one}), which determines the first-order
Volterra term $\gamma_1(t)$, in general contains time-dependent coefficients.

Because the Volterra series generalizes the Taylor-Maclaurin expansion of functions
in calculus \cite[Sec. 1.5]{Rugh}, Eq.~(\ref{eq:Volterra}) may be restricted by a
radius of convergence or may even fail to exist for some choices of $x_0(t)$. The
equilibrium point $x_0(t) \equiv 0$, which is a convenient choice for the monostable
case of Eq.~(\ref{eq:main}), is unstable in the regime of relaxation oscillations
and yields a divergent kernel $g_1(t)$. With $x_0(t)\equiv0$ we therefore cannot
construct an approximate representation~(\ref{eq:xone}) that is valid for long time
scales \cite{Belousov2019I}.

For the above reason we use the Volterra series expansion about $x_0(t)$ that represents
the stable limit-cycle solution of Eq.~(\ref{eq:zero}). Because a closed-form expression
of this solution is unknown, as its approximation one may adopt a truncated Fourier
expansion $x_0(t)\approx\xi(t)$ that can be obtained by various methods~\cite[Sections
4.4 and 5.9]{JordanSmith}. Substituting $\xi(t)$ for $x_0(t)$ in Eq.~\eqref{eq:one}
we obtain a linear problem
\begin{equation}\label{eq:floquet}
  \ddot\gamma_1 + a_\xi \dot\gamma_1 + b_\xi \gamma_1 = f,
\end{equation}
with time-dependent periodic coefficients
\begin{align}
  &a_\xi(t) = a+c \xi(t)^2,\\
  &b_\xi(t) = b + 2 c \dot\xi(t)\xi(t).
\end{align}

Note that the time-dependent friction $a_\xi(t)$ and stiffness $b_\xi(t)$ oscillate
around positive average values that ensure the stability of Eq.~(\ref{eq:floquet}).
These coefficients are statistically independent from the driving white-noise force $f(t)$
at all times. On average the response of the linear stochastic Eq.~(\ref{eq:floquet})
can be therefore described by effective friction and stiffness constants. To implement
this simplification for the quasiperiodic term $\gamma_1(t)$, in the spirit of time-averaging
methods \cite[Chapter 4 in ][Sec. 9.2]{JordanSmith,Grimshaw2017} we replace the periodic
coefficients in Eq.~(\ref{eq:floquet}) by their mean values
\begin{align}\label{eq:avga}
  &\avg{a_\xi(t)} = \int_0^{2\pi / \sqrt{b}} \frac{\sqrt{b} dt}{2\pi}a_\xi(t),
  \\\label{eq:avgb}
  &\avg{b_\xi(t)} = \int_0^{2\pi / \sqrt{b}} \frac{\sqrt{b} dt}{2\pi}b_\xi(t),
\end{align}
in which the ensemble average of a periodic function is related to the time average
over one period $2\pi/\sqrt{b}$. In this approximation Eq.~(\ref{eq:floquet}) describes
a harmonic oscillator $\tilde\gamma(t) \approx \gamma_1(t)$:
\begin{equation}\label{eq:ho}
  \ddot{\tilde\gamma} - \avg{a_\xi}\dot{\tilde\gamma} + \avg{b_\xi} \tilde\gamma = f
\end{equation}
with the linear response function
\begin{equation}\label{eq:gxi}
  g_\xi(t) = \Omega^{-1} \me^{-\frac{\avg{a_\xi} t}{2}} \sin(\Omega t) \approx g_1(t),
\end{equation}
in which $\Omega = \sqrt{\avg{b_\xi} - \avg{a_\xi}^2/4}$. If $\Omega^2 < 0$ one
should use $\sqrt{\avg{a_\xi}^2/4 - \avg{b_\xi}}$ instead of $\Omega$ and replace
the trigonometric sine in Eq.~(\ref{eq:gxi}) by the hyperbolic one \cite[Sec. II-3]{BelousovCohen,Chandrasekhar}.

The approximate solution of the stochastic Van der Pol equation~(\ref{eq:main}) is
thus expressed by a sum of two \textit{independent} contributions $\xi(t)$ and $\tilde\gamma(t)$
\begin{equation}\label{eq:solution}
  x(t) \approx x_\xi(t) = \xi(t) + \tilde\gamma(t) = \xi(t) + \int_0^t ds g_\xi(t-s) f(s).
\end{equation}
The linear-response term $\tilde\gamma(t)$ in the above equation has a Gaussian probability
density, with a zero mean $\avg{\tilde\gamma}=0$ and an autocovariance function
\cite{BelousovCohen}
\begin{multline}\label{eq:covgamma}
  \avg{\tilde\gamma(0)\tilde\gamma(t)} = \frac{A^2}{2 \avg{a_\xi} \avg{b_\xi}}
  \exp\left(-\frac{\avg{a_\xi} t}{2}\right)
    \\\times\left[\cos(\Omega t) + \frac{\avg{a_\xi}}{2\Omega}\sin(\Omega t)\right].
\end{multline}

\begin{figure*}[t]
\includegraphics[width=1\textwidth]{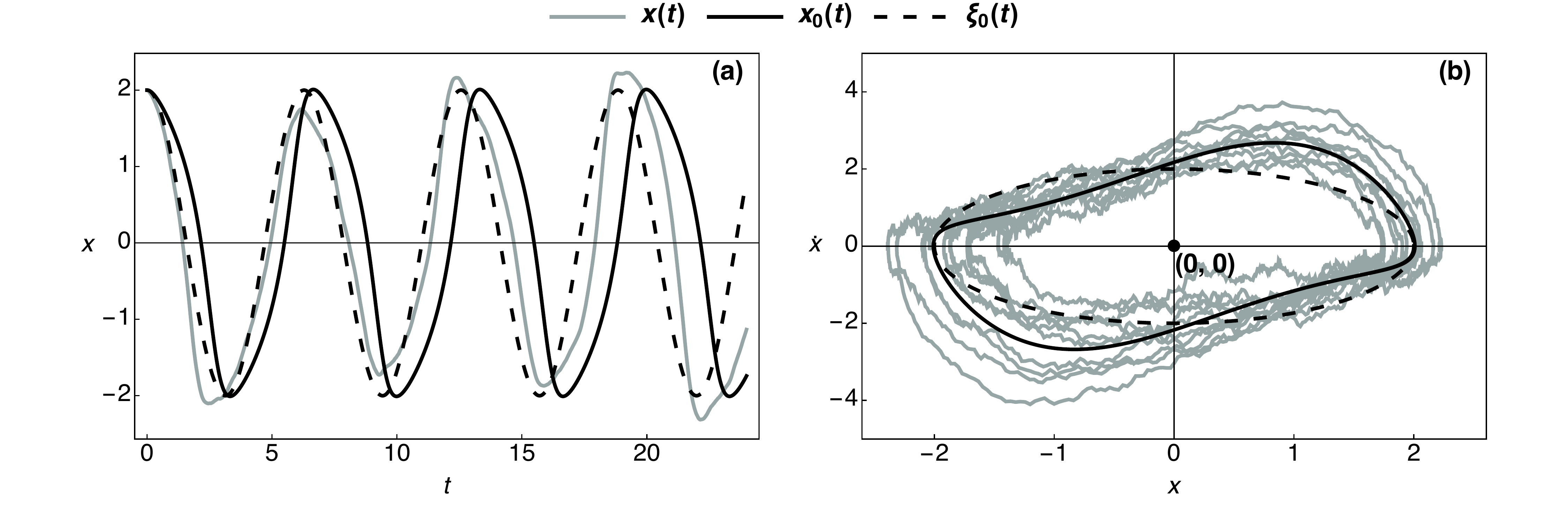}
\caption{\label{fig:zero}
  Comparison of the noisy Van der Pol oscillator $x(t)$ with the limit-cycle solution
  $x_0(t)$ and an approximate expression $\xi_0(t)$. The system parameters are $\mu=1$
  and $A=0.6$ in the reduced units (Sec.~\ref{sec:num}), whereas the initial conditions
  are set to $(x,\,\dot{x})=(\alpha,\,0)$. Panel (a): phases of the noisy oscillations'
  peaks fluctuate around the maxima of $x_0(t)$; the theoretical expression $\xi_0(t)$
  [Eq.~\eqref{eq:xi0}] captures the overall trend of the time series $x(t)$. Panel~(b):
  comparison of the phase-space orbits $\big{(}x(t),\,\dot{x}(t)\big{)}$,
  $\big{(}x_0(t),\,\dot{x}_0(t)\big{)}$, and $\big{(}\xi_0(t),\,\dot{\xi_0}(t)\big{)}$;
  the deviations of the Van der Pol limit cycle $x_0(t)$ from the single-mode harmonic
  approximation $\xi_0(t)$ is less than or comparable to the uncertainty of the trajectory
  $x(t)$.
}
\end{figure*}

In Appendix~\ref{app:sol} we derive two levels of approximations for the autonomous
term $x_0(t)$, \textit{viz.} $\xi_0(t)$ and $\xi_1(t)$ [Eqs.~\eqref{eq:xi0} and \eqref{eq:xi1}].
A comparison of the noisy Van der Pol oscillator $x(t)$ with the limit-cycle solution
$x_0(t)$ and a single-mode Fourier expansion $\xi_0(t)$ is shown in Fig.~\ref{fig:zero}.
The trajectory $\xi_0(t)$ approximates well the period of oscillations and the overall
trend of the time series $x(t)$. For moderate values of the noise amplitude $A$ and
the parameter $\mu=-a/\sqrt{b}$, which controls the nonlinear character of oscillations
(Sec.~\ref{sec:num}), the error of the single-mode approximation $x_0(t)\approx\xi_0(t)$
is less than or comparable to the uncertainty of the trajectory $x(t)$.

The approximate solution $x_\xi(t)$ is limited to small noise amplitudes not only
due to the truncation error in Eq.~(\ref{eq:Volterra}). The external stochastic force
induces variations of the oscillator's phase $\phi=\sqrt{b}t_0$ in the periodic term
$\xi(t)\to\xi(t-t_0)$. This effect of noise is especially large when the system's
trajectory is driven close to the point $(x,\dot{x}) = (0,0)$. Crossing this point
may cause a shift through a phase angle as large as $\phi=\pi$, which requires in
general a large force $f(t)$.

At moderate noise amplitudes the above phase variations average to zero
[Fig.~\ref{fig:zero}(a)]. Consequently, the time-invariant statistics of $x_\xi(t)$,
which is analyzed in Appendix~\ref{app:sta}, agrees with that of the noisy Van der
Pol oscillator $x(t)$. Their time autocorrelation functions coincide, however, only
at small time $t$ (Appendix~\ref{app:sta}, Fig.~\ref{fig:chi}).

Finally, we remark that the approximate solution Eq.~\eqref{eq:solution} can be generated
by a forced harmonic oscillator. Such a representation provides a way to emulate
self-sustained oscillations of the Van der Pol type by using a linear system with
a periodic driving, which is much simpler to analyze quantitatively. This idea is
demonstrated in Sec.~\ref{sec:exp}.

\subsection{\label{sub:pinf} Parametric inference}
Equation~(\ref{eq:solution}), together with the analysis presented in Appendices~\ref{app:sol}
and \ref{app:sta}, encompasses a simple inference technique that can be used to extract
the parameter values $a$, $b$, $c$, and $A$ for Eq.~\eqref{eq:main} directly from
time series of the Van der Pol oscillator. The procedure consists of two curve-fitting
steps. First we determine the parameters $a$ and $b$ from the empirical autocorrelation
function of the time series $x(t)$. Then we extract the parameters $c$ and $A$ from
the oscillatory trend of the trajectory and the variance $\avg{x^2}$, respectively.

To fit the empirical autocorrelation function $\chi(t)$ we adopt a theoretical Eq.~\eqref{eq:chi1}
from Appendix~\ref{app:sta} in the form
\begin{widetext}
\begin{multline}\label{eq:chi}
  \chi(t) \simeq
    \frac{\lambda_1^2 \cos(\omega t)}{(\lambda_1^2 + \lambda_2^2)(1 + 5\mu^2/32)}
      \left\{
        1 + \frac{\mu^2}{32}[4 + \cos(2\omega t)]
      \right\}
      + \frac{\lambda_2^2}{\lambda_1^2 + \lambda_2^2}
      \me^{-\left(1 + \frac{5\mu^2}{16}\right)\frac{\mu\omega t}{2}}
      \left\{
        \cos\left[\frac{\omega t}{2} \sqrt{4-\left(1 + \frac{5\mu^2}{16}\right)^2 \mu^2} \right]
        \right.\\\left.
        + \mu \left(1 + \frac{5\mu^2}{16}\right)
        \left[4-\left(1 + \frac{5\mu^2}{16}\right)^2 \mu^2\right]^{-1/2}
        \sin\left[\sqrt{4-\left(1 + \frac{5\mu^2}{16}\right)^2 \mu^2} \frac{\omega t}{2}\right]
      \right\}.
\end{multline}
\end{widetext}
Because the fitting constants $\lambda_1^2\propto\avg{\xi^2}$ and
$\lambda_2^2\propto\avg{\tilde\gamma^2}$ are determined up to an arbitrary factor,
they are treated as nuisance parameters in the above expression.

Equation~\eqref{eq:chi} approaches its first zero as $t \to \tau \approx \pi/(2\sqrt{b})$,
approximately a quarter period of the trigonometric factors in this theoretical expression.
We may then apply the criterion of Lagarkov and Sergeev \cite{Lagarkov1978,BelousovCohen}
to select the interval $0 \le t \le \tau$ over which Eq.~\eqref{eq:chi} is expected
to be accurate, that is, the initial decay of the empirical time autocorrelations
(Appendix~\ref{app:sta}). Because this theoretical expression is very flexible, the
initial guess of the fitting constants must be chosen with care. For the best performance
we suggest using $\mu\lesssim1$, $\omega\sim\pi/(2\tau)$, $\lambda_{1,2} \sim \sqrt{\avg{x^2}/2}$.
The parameters of interest $a = -\mu \omega$ and $b=\omega^2$ are then found from
the optimized values of the constants $\mu$ and $\omega$. To estimate the uncertainties
of $\mu$ and $\omega$ we repeat the fitting procedure over few slightly longer intervals
of duration $\tau < t < 2 \tau$.

In the next step of the inference method we estimate the amplitude of the Van der Pol
limit-cycle oscillations. Equation~\eqref{eq:solution} decomposes the trajectory
$x(t)$ into a sum of the oscillatory term $\xi_0(t)\approx\xi(t)$, that determines
the average trend, and the Gaussian random-error term $\tilde\gamma(t)$. As discussed
in Sec.~\ref{sub:lrsp}, the limit-cycle solution $x_0(t)$ does not account for the
slowly fluctuating phase of the noisy Van der Pol oscillations. As in the case of
the stochastic Duffing oscillator \cite{Belousov2019I}, we circumvent this issue
by applying Eq.~\eqref{eq:solution} \textit{locally}: the time series of $x(t)$ can
be split into pieces $x_+(t)$ and $x_-(t)$ for, respectively, $x(t) > 0$ and $x(t)<0$.
The duration of each component corresponds approximately to a half period $\pi/\sqrt{b}$
of $\xi(t)$. Assuming that the phase shift is constant over one period of oscillations,
we then fit these pieces of the whole trajectory to the following formula:
\begin{equation}\label{eq:xi}
  x_0(t) \simeq \xi_0(t - t_0) =
    \alpha_c \cos(\sqrt{b}t) + \alpha_s \sin(\sqrt{b}t),
\end{equation}
in which
\begin{equation}
  \alpha_c = \alpha \cos(\sqrt{b} t_0),\quad
  \alpha_s = \alpha \sin(\sqrt{b} t_0)
\end{equation}
\textit{cf. } Eq.~\eqref{eq:xi0} in Appendix~\ref{app:sol}. Note that the constant
$b$ in Eq.~\eqref{eq:xi} is fixed to the value estimated from the first step of
the method. We also ensure that fitted trajectories have a minimal duration of $\pi b^{-1/2}/2$.

From the optimized values of the fitting constants $\alpha_c$ and $\alpha_s$, we
obtain the amplitude of limit-cycle oscillations and the remaining parameters of
interest:
\begin{equation}\label{eq:alphacA}
  \alpha = \sqrt{\alpha_c^2 + \alpha_s^2},\;
  c = -\frac{4 a}{\alpha^2},\;
  A = \sqrt{a b (2 \avg{x^2} - \alpha^2)},
\end{equation}
in which $\avg{x^2}$ is the sample variance of the empirical time series $x(t)$.
The parameter $\alpha$ and its uncertainty are determined by averaging over all trajectory
pieces $x_\pm(t)$.

The numerical error of fitting the approximate Eq.~\eqref{eq:xi} to the trajectory
pieces $x_\pm(t)$ eventually may exceed the uncertainty of the driving noise $f(t)$.
Therefore, when the autonomous term $\xi_0(t)\propto\alpha$ dominates the statistical
variability of the data, a small noise amplitude $A\to0$ cannot be inferred accurately.
Unfortunately a more elaborate approximation $\xi(t)\simeq\xi_1(t-t_0)$ [Appendix~\ref{app:sol},
Eq.~\eqref{eq:xi1}] cannot address this issue. As Fourier series are able to match
almost any curve arbitrarily close with a sufficient number of terms, the truncated
higher-order expansion $\xi_1(t)$ overfits noisy trajectories of $x(t)$.

\section{\label{sec:num}Application to simulated data}
In the system of units reduced by a time constant $b^{-1/2}$ and a length constant
$\sqrt{-a/c}$, Eq.~(\ref{eq:main}) takes a canonical form \cite[Sec. 7.4 and 7.5]{StrogatzII}
\begin{equation}\label{eq:canon}
  \ddot{x} - \mu (1 - x^2)\dot{x} + x = \frac{A}{b\sqrt{-a/c}}\dot{w},
\end{equation}
in which the parameter $\mu = -a/\sqrt{b}$ controls the nonlinear character of the
dynamics. The greater its value, the larger is the amplitude of the relaxation oscillations.
This parameter represents the ratio of two time scales $b^{-1/2}$ and $-a = \mu b^{-1/2}$.

Two control parameters of Eq.~(\ref{eq:canon}) that are not fixed in the system of
reduced units are $\mu$ and $A$. Without external driving the Van der Pol oscillator,
which orbits around the origin of the phase space with the amplitude $\alpha = 2 \sqrt{-a/c}$
in the harmonic potential $U(x) = b x^2/2$ [Fig.~\ref{fig:zero}], has an energy scale
$$ U(\alpha) = b \alpha^2/2 = 2 \mu b^{3/2}/c. $$
The energy scale of the external force $f(t) = A \dot{w}(t)$ is $A^2/\sqrt{b}$. One
might therefore expect the small-force expansion Eq.~(\ref{eq:xone}) to hold for
$$ A < \sqrt{\sqrt{b} U(\alpha)} = b \sqrt{2 \mu/c}, $$
which relates the two control parameters of Eq.~(\ref{eq:canon}) $A$ and $\mu$.

To test the theory presented in the previous section, we simulated Eq.~(\ref{eq:main})
for selected values of noise amplitudes $A$. The computational details are summarized
in Appendix~\ref{app:sim}. As a typical value we choose to fix the parameter $\mu = 1$.
For increasingly large values of $\mu$ Eq.~(\ref{eq:xi}) becomes progressively less
accurate. The techniques that we propose in the present paper should work also for
$\mu > 1$, but their precision deteriorates for larger values of this parameter.
We discuss the accuracy of the derived theoretical expression in full detail in
Appendices~\ref{app:sol} and \ref{app:sta}.

The efficiency of the parametric-inference method that we described in Sec.~\ref{sub:pinf}
is demonstrated by Table~\ref{tbl:fit}. Our approach renders best estimates of the
model parameters at a moderate level of noise $0.2 < A < 1.2$. On one side, the truncation error of Eq.~\eqref{eq:Volterra} grows with the amplitude $A$ as the nonlinear effects become increasingly important. On the other side, because
relaxation oscillations have nontrivial statistics even in the absence of external
forces, it is difficult to discriminate between the numerical errors of fitting approximate
expressions and the stochastic uncertainty of the driving noise.

\begin{table}[t]
\caption{\label{tbl:fit}
  Inference of the Van der Pol oscillator's parameter values from time series of
  Eq.~(\ref{eq:main}) that were simulated with $a=-1$, $b=1$, $c=1$ fixed and $A$
  varied in the range $[0,1.2]$. The estimated parameter values are denoted by $\hat{a}$,
  $\hat{b}$, $\hat{c}$, and $\hat{A}$, respectively.
}
\begin{ruledtabular}
\begin{tabular}{c | r r r r}
$\hat{A}$ & $\hat{a}$ & $\hat{b}$ & $\hat{c}$ & $\hat{A}$ \\
\hline
$0.0$ & $-0.97\pm0.01$ & $0.891\pm0.003$  & $0.94\pm0.01$ & $0.218\pm0.002$ \\
$0.2$ & $-0.97\pm0.02$ & $0.890\pm0.004$  & $0.95\pm0.02$ & $0.26\pm0.04$ \\
$0.4$ & $-0.98\pm0.02$ & $0.899\pm0.004$  & $0.97\pm0.02$ & $0.36\pm0.05$ \\
$0.6$ & $-1.01\pm0.02$ & $0.912\pm0.004$  & $1.02\pm0.03$ & $0.55\pm0.06$ \\
$0.8$ & $-1.02\pm0.03$ & $0.929\pm0.005$  & $1.07\pm0.04$ & $0.76\pm0.06$ \\
$1.0$ & $-1.04\pm0.03$ & $0.941\pm0.007$  & $1.12\pm0.05$ & $0.97\pm0.06$ \\
$1.2$ & $-1.06\pm0.03$ & $0.96\pm0.01$  & $1.17\pm0.04$ & $1.16\pm0.06$ \\
\end{tabular}
\end{ruledtabular}
\end{table}

Except for the marginal cases of small and large noise amplitudes $A$, the values
of all parameters in Table~\ref{tbl:fit} are accurate within $15\%$. The constant
$b$ is determined with the largest bias, because our theoretical expressions overestimate
the frequency of the Van der Pol limit-cycle oscillations [Figs.~\ref{fig:zero}(a)
and \ref{fig:one}(a)]. We remark that the value of the parameter $\mu$ controlling
the nonlinear character of the dynamics is estimated within a ten-percent error.

\section{\label{sec:exp}Application to experimental data}
In this section, we apply the theory of Sec.~\ref{sec:thr} to experimental data for
a real physical system. We consider an example from biology. Various models with
a limit-cycle behavior have been proposed to describe self-sustained oscillations
of a hair bundle---a mechanosensitive organelle of the receptor cells in the inner
ear of vertebrates. The spontaneous undulatory motion of the hair bundle's position
has been related to an active process in the ear that amplifies acoustic signals,
sharpens frequency selectivity, and broadens the operational dynamic range \cite{Hudspeth2014}.
Much theoretical and experimental effort has been devoted to understand the origin
of these oscillations and their behavior
\cite{Hudspeth2014,martin1999,faber2018,Nadrowski2004,Tinevez2007,Barral2010,ONH2012}.

To apply our theory to time series of a hair-bundle's oscillation, we recorded movement
of a hair cell bundle as described previously in
Refs.~\cite{martin1999,azimzadeh2017,azimzadeh2018}. We directly projected a high-contrast
image of the tip of an oscillating hair bundle onto a dual photodiode and recorded
its calibrated movement as a function of time \cite{foot1}. Recently proposed models
of these oscillations \cite{Nadrowski2004,Tinevez2007,Barral2010,ONH2012} can be
explicitly related to the class of Lienard systems. Among others, the simple Van
der Pol Eq.~(\ref{eq:main}) has also been considered to describe the active process
in hearing organs \cite{Gelfand2010}.

Using our theoretical approach, we address two problems of modeling active oscillations
of a hair bundle. First, can the simple Van der Pol Eq.~(\ref{eq:main}) explain
experimental time series of these oscillations? And, if not, is it possible to
relate the hair-cell bundle oscillations to the Van der Pol equation in a more general
setting?

\begin{figure*}[t]
\includegraphics[width=1\textwidth]{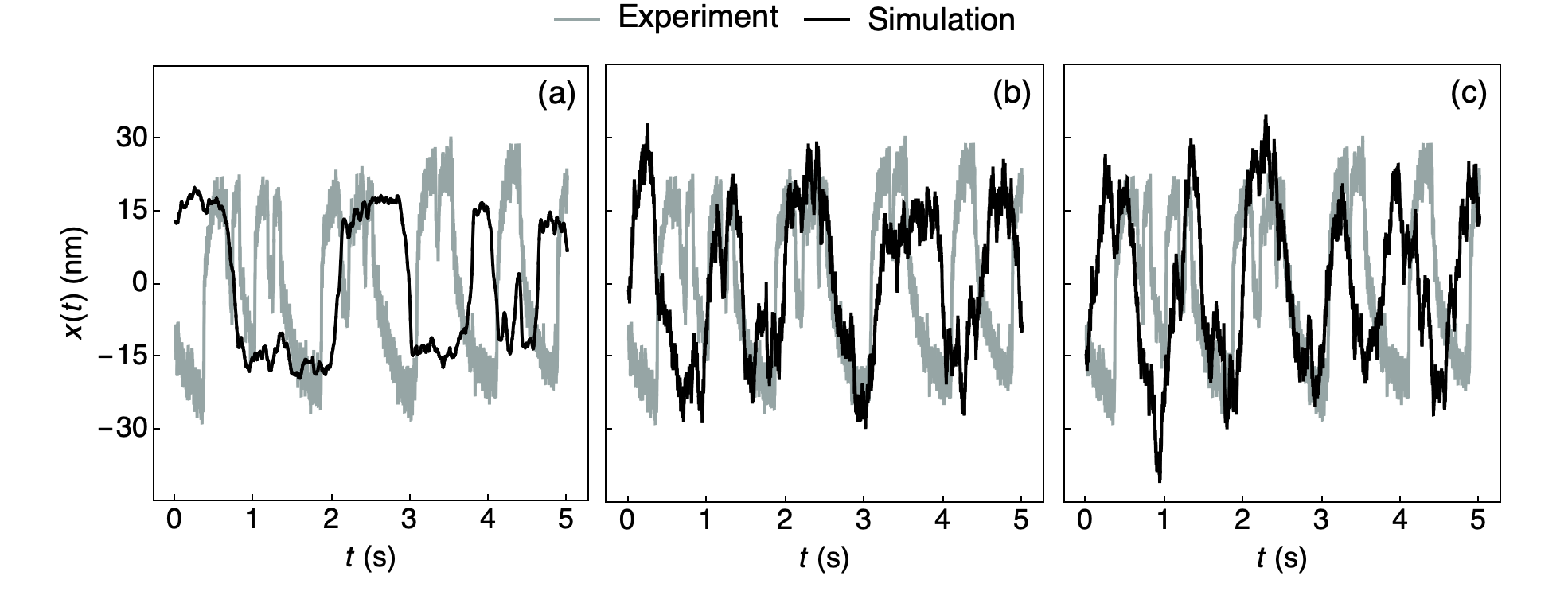}
\caption{\label{fig:exp}
  Comparision of three models for hair-cell bundle oscillations with the experimental
  data. Panel (a): The simple Van der Pol Eq.~\eqref{eq:main} does not reproduce
  oscillatory features of the experimental measurements. Panel (b): The hidden Van
  der Pol Eqs.~\eqref{eq:xz}--\eqref{eq:ddz} capture the general character of the
  hair bundle's oscillations. Panel (c): The effective linear Eqs.~\eqref{eq:xz}
  and \eqref{eq:zho} generally agree with the experimental data, but overestimate
  slightly the peak values of hair bundle's noisy oscillations.
}
\end{figure*}

Our answer to the first of the two questions is negative. If we suppose that our
experimental observations of $x(t)$ come from Van der Pol oscillator, then the
method of Sec.~\ref{sub:pinf} is applicable to our data directly as presented.
A simulation of Eq.~\eqref{eq:main} with the parameter values obtained by these means
is compared with our experimental data in Fig.~\ref{fig:exp}(a). Evidently the time
series of a simple Van der Pol model do not resemble the oscillations of a hair-cell bundle.

By using few simplifying assumptions, the alternative models mentioned above
\cite{Nadrowski2004,Tinevez2007,Barral2010,ONH2012} can be reduced to a form that
is directly related to Eq.~(\ref{eq:main}). A convenient scheme is a linear coupling
between the coordinate $x(t)$ and a hidden Van der Pol oscillator $z(t)$. For a detailed
demonstration of our approach we choose a simple scheme that is based on a parsimonious
model of Ref.~\cite{ONH2012}:
\begin{align}\label{eq:xz}
  &\dot{x} = \dot{z} + c_x z,
  \\\label{eq:ddz}
  &\ddot{z} + a \dot{z} + b z + c \dot{z} z^2 = A \ddot{w},
\end{align}
in which $c_x$ is the coupling constant, whereas $a$, $b$, and $c$ are analogous
to Eq.~\eqref{eq:main}. Note the double overdot on the right-hand side of Eq.~\eqref{eq:ddz}.
We spare the mathematical and numerical details of Eqs.~\eqref{eq:xz} and \eqref{eq:ddz}
for Appendix~\ref{app:hcb}.

Extension of the theory presented in Sec.~\ref{sec:thr} is quite straightforward.
Because our example regards a drastic approximation of the original system, for simplicity
we provide below only the formulas derived from the zeroth-order approximation of
the Van der Pol limit cycle [Eq.~\eqref{eq:xi0}]. For the hair bundle's position
we obtain an equation analogous to (\ref{eq:solution}):
\begin{equation}\label{eq:zeta}
  x(t) \simeq \zeta(t) + \tilde\zeta(t),
\end{equation}
in which $\zeta(t)$ and $\tilde\zeta(t)$ are the autonomous and the linear-response
terms analogous to $\xi(t)$ and $\tilde\gamma(t)$. Instead of Eq.~\eqref{eq:xi} we
obtain from \eqref{eq:xz} and \eqref{eq:xi0}
\begin{multline}
  \zeta(t-t_0) = \alpha \cos[\sqrt{b} (t-t_0)] + \alpha c_x \int_0^{t-t_0} ds \cos(\sqrt{b} s)
    \\ = \tilde\alpha_c \cos(\sqrt{b} t) + \tilde\alpha_s \sin(\sqrt{b} t),
\end{multline}
in which $\tilde\alpha_c = \alpha_c - \kappa \alpha_s$, $\tilde\alpha_s = \alpha_s + \kappa \alpha_c$,
$\kappa = c_x/\sqrt{b}$; and instead of Eq.~\eqref{eq:chi} we get
\begin{multline}
  \tilde\chi(t)  =
    \frac{\lambda_1^2}{\lambda_1^2+\lambda_2^2} \cos(\omega t)
      + \frac{\lambda_2^2}{\lambda_1^2+\lambda_2^2}
      \\\times \exp\left(-\frac{\mu \omega t}{2}\right)\left[
        \cos\left(\sqrt{4 - \mu^2}\frac{\omega t}{2}\right)
      \right.\\\left.
      + \frac{\mu (\kappa^2-1)}{\sqrt{4-\mu^2}(\kappa^2+1)}
        \sin\left(\sqrt{4 - \mu^2}\frac{\omega t}{2}\right)
      \right].
\end{multline}
In addition we must replace the expression for $\alpha$ and $A$ in Eq.~\eqref{eq:alphacA}
by
\begin{eqnarray}
  \alpha &=& \sqrt{\frac{\tilde\alpha_c^2 + \tilde\alpha_s^2}{(1+\kappa^2)}},\\
  A &=& \sqrt{-a \left(\frac{2 \avg{x^2}}{1 + \kappa^2} - \alpha^2\right)}.
\end{eqnarray}

By fitting the empirical autocorrelations and the oscillatory trend of the experimental
measurements with the above formulas, we can infer all the parameter values for
Eqs.~\eqref{eq:xz} and \eqref{eq:ddz}. As illustrated in Fig.~\ref{fig:exp}(b), these
dynamical equations reproduce closely the character of the hair bundle's oscillations
and their frequency, despite the strong assumptions used to simplify the original
model of Ref.~\cite{ONH2012}.

Finally, as anticipated in Sec.~\ref{sub:lrsp}, we present below an effective linear
model that imitates the self-sustained oscillations generated by the nonlinear system
of Eqs.~\eqref{eq:xz} and \eqref{eq:ddz}. One may recognize that the Gaussian term
$\tilde\zeta(t)$ in Eq.~\eqref{eq:zeta} [as well as $\tilde\gamma(t)$ in Eq.~\eqref{eq:solution}]
represents a harmonic oscillator driven by a white-noise signal. If we apply to this
oscillator a specifically designed deterministic force, in addition to the stochastic
fluctuations we can elicit a response composed of the same Fourier modes that are
present in the term $\zeta(t)$ [or $\xi(t)$].

The above program is implemented by coupling Eq.~\eqref{eq:xz} to \eqref{eq:zho} that
is derived in Appendix~\ref{app:hcb}. The exact steady-state solution of this linear
system, whose simulation is compared with the experimental data in Fig.~\ref{fig:exp}(c),
is then given by Equation~\eqref{eq:zeta}. The time series of the dynamical Eqs.~\eqref{eq:xz}
and \eqref{eq:zho} are nearly indistinguishable from the original system
\eqref{eq:xz}--\eqref{eq:ddz}.

\begin{figure}[t]
\includegraphics[width=1\columnwidth]{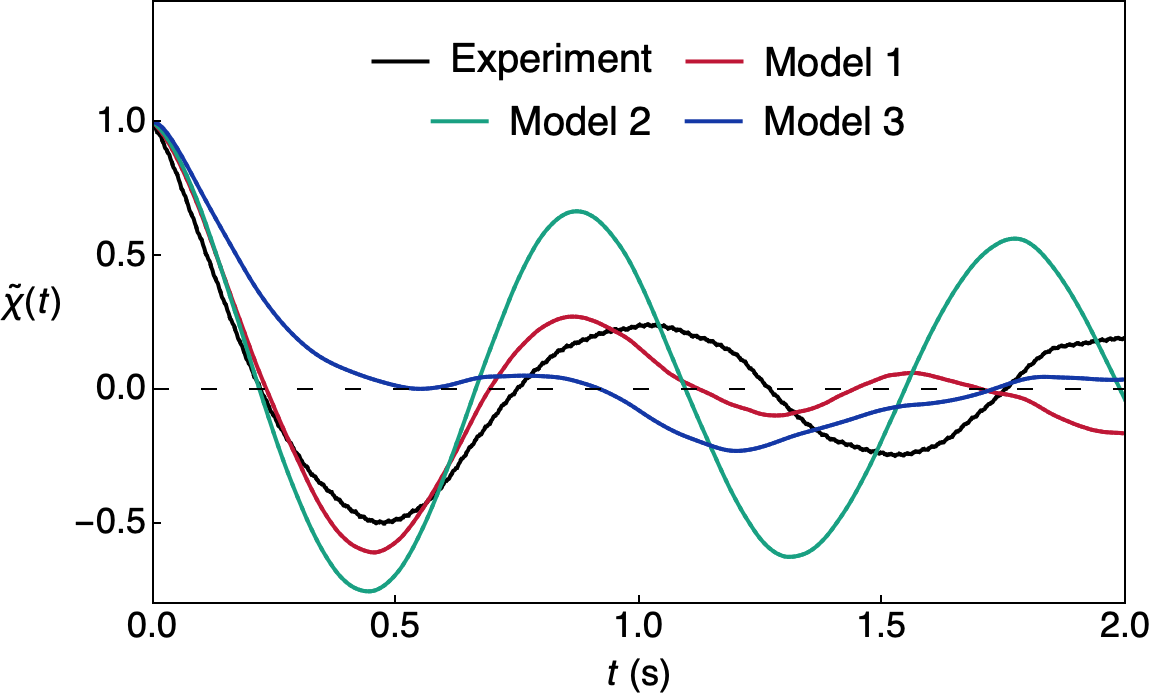}
\caption{\label{fig:auto}
  Comparison of the empirical time autocorrelation function $\tilde\chi(t)$ between
  the expeirments and the three theoretical models:
  1 -- the hidden Van der Pol Eqs.~\eqref{eq:xz} and \eqref{eq:ddz},
  2 -- the effective linear Eqs.~~\eqref{eq:xz} and \eqref{eq:zho},
  3 -- the simple Van der Pol Eq.~\eqref{eq:main}.
}
\end{figure}

The empirical time autocorrelation functions of the experimental system and the three
models discussed above are compared in Fig.~\ref{fig:auto}. The Van der Pol oscillator
does not match the observations at all, whereas the system with the hidden Van der
Pol oscillator and its linear imitation reproduce the oscillatory features of the
hair bundle movements quite well.

More advanced models of the hair-cell bundle oscillations can also be analyzed with
help of the methods proposed in this paper. These developments, which require additional
mathematical details, will be a subject of our future communications.

\section{Conclusion}

Using the Volterra series we have analyzed statistical features of a noisy Van der
Pol equation. Perhaps surprisingly, its solution can be decomposed within the linear
order of the driving force into two independent contributions: a deterministic part
that describes relaxation oscillations, and a stochastic linear-response term. With
the help of simple approximation schemes we showed that the deterministic contribution
has a singular probability density, whereas the stochastic part can be described
by a Gaussian process with a second-order autocorrelation function.

Volterra series provide a representation of solutions for nonlinear stochastic equations.
Other theoretical approaches, such as the Fokker-Planck equation and path integrals,
focus instead on statistical properties of an ensemble of systems' realizations and
offer less information about their dynamics. The theoretical tools may complement
each other; for instance, the Volterra series may be used to an advantage in ergodic
problems when time averaging is more convenient than ensemble averaging for the evaluation
of statistical properties.

The inference method based on our analytical results allows us to estimate parameter
values of the stochastic Van der Pol model from observed time series of oscillations
for moderate levels of the driving noise. However, due to the approximate nature of
our theoretical expressions, this method cannot determine accurately values of small
noise amplitudes. Two problems pose the major challenge for the Volterra-series approach
here: finding a faithful representation of the Van der Pol limit-cycle solution and
modeling the fluctuating phase of noisy oscillations. The latter issue is perhaps
more pressing. A viable approach to the problem of fluctuating phase could be to
study Eq.~\eqref{eq:main} in polar coordinates \cite{Raphael1993}.

In a simplified case study we have demonstrated that our theory can be applied to
analyze actual physical systems. In particular, the Volterra-series approach offers
a method of constructing a linear model that imitates the dynamics of self-sustatined
oscillations. Albeit approximate, this imitation can be used to simplify quantitative
studies of complex systems.

\begin{acknowledgments}
We thank Andrew R. Milewski and B. Fabella for their assistance with the experiments.
\end{acknowledgments}

\section*{Appendices}
\appendix

\section{\label{app:sol} Limit-cycle solution of the autonomous Van der Pol oscillator}
In this appendix we derive two levels of approximation for the limit-cycle solution
of the autonomous Van der Pol problem. Although these two expressions can be obtained
by using the harmonic-balance and Lindstedt-Poincare methods, we adopt here a unifying
variational Green's-function approach, which is similar in spirit to that of
Refs.~\cite{Khuri2017,Abukhaled2017}. Equation~(\ref{eq:zero}) that we are
solving can be recast as
\begin{equation}\label{eq:Lzero}
  \oper{L}_\omega x_0 = - a \dot{x}_0 - (b-\omega^2) x_0 - c \dot{x}_0 x_0^2,
\end{equation}
in which $\oper{L}_\omega = \partial_t^2 + \omega^2$ is a linear differential operator
with a constant frequency parameter $\omega > 0$. As in the harmonic-balance and
Lindstedt-Poincare methods \cite[Sec. 4.4 and 5.9]{JordanSmith}, we use the initial
condition $\Big{(}x_0(0),\,\dot{x}_0(0)\Big{)} = (\alpha,\,0)$ with $\alpha$ left
unspecified. The solution of Eq.~(\ref{eq:Lzero}) must then satisfy
\begin{align}\label{eq:x0}
  x_0(t) =& \xi_0(t) - \int_0^t g_\omega(t-s)
    \nonumber\\
    &\times [
      a \dot{x}_0(s) + (b-\omega^2) x_0(s) + c\dot{x}_0(s) x_0^2(s)
    ]
    \nonumber\\
  =& \alpha \cos(\omega t)
    -\frac{\sin (\omega t)}{\omega} \int_0^t \cos(\omega s)
    \nonumber\\
    &\times [
      a \dot{x}_0(s) + (b-\omega^2) x_0(s) + c\dot{x}_0(s) x_0^2(s)
    ]
    \nonumber\\
    &+\frac{\cos (\omega t)}{\omega} \int_0^t \sin(\omega s)
    \nonumber\\
    &\times [
      a \dot{x}_0(s) + (b-\omega^2) x_0(s) + c\dot{x}_0(s) x_0^2(s)
    ],
\end{align}
in which $\xi_0(t) = \alpha \cos(\omega t)$ solves the equation $\oper{L}_\omega \xi_0 = 0$,
whereas $g_\omega(t) = \sin(\omega t)/\omega$ is the Green function associated with
the operator $\oper{L}_\omega$.

In the first approximation we posit a single-mode Fourier expansion $x_0(t) \approx \xi_0(t)$.
For the right-hand side of Eq.~\eqref{eq:x0} to satisfy the periodic boundary condition
$x(0)=x(2\pi/\omega)$, one must choose:
\begin{equation}\label{eq:xi0}
  \xi_0(t) = \alpha \cos(\sqrt{b}t)
\end{equation}
with $\alpha = 2\sqrt{-a/c}$. Alternatively this solution can be obtained by the method
of harmonic balance \cite[Sec. 4.4]{JordanSmith}.

The single-mode solution $\xi_0(t)$ can be further improved by one Picard iteration:
we substitute $\xi_0(t)$ for $x_0(t)$ on the right-hand side of Eq.~(\ref{eq:x0})
and complete the integration to get
\begin{equation}\label{eq:xi1}
  \xi_1(t) = \alpha \left[
    \cos(\sqrt{b} t)
    + \frac{3 \mu}{8}\sin(\sqrt{b} t)
    - \frac{\mu}{8}\sin(3\sqrt{b} t) \right
  ].
\end{equation}
The above expression coincides with the perturbative solution that can be obtained
by the Lindstedt-Poincare method of two time scales within the linear order of the
parameter $\mu=-a/\sqrt{b}$ (Sec.~\ref{sec:num}). With respect to this parameter,
Eq.~\eqref{eq:x0} represents the zeroth-order approximation of the limit cycle.

\begin{figure*}[t]
\includegraphics[width=1\textwidth]{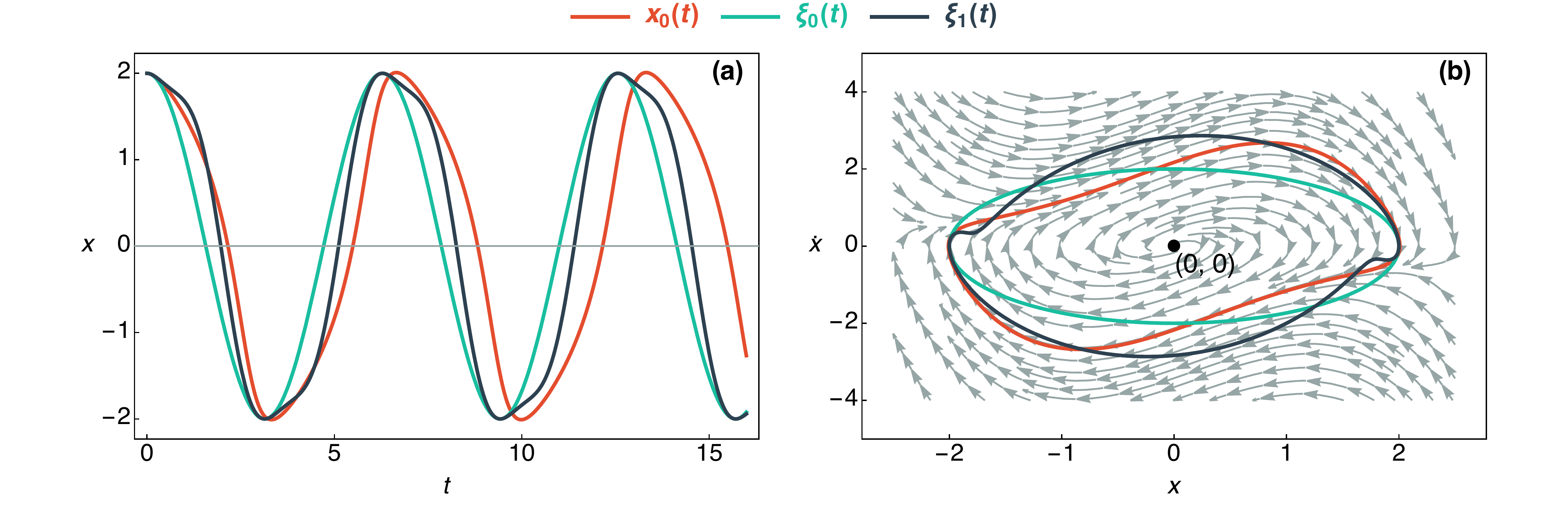}
\caption{\label{fig:one}
  Comparison of the approximate solution $\xi_0(t)$ and $\xi_1(t)$ given by Eqs.~(\ref{eq:xi0})
  and (\ref{eq:xi1}), respectively, with a simulation of the autonomous Van der Pol
  Eq.~(\ref{eq:zero}):
  (a) time series of the limit-cycle solution
  (b) orbit of the oscillator's limit cycle in the phase space $(x_0,\,\dot{x}_0)$.
  The simulation parameters are $\mu=1$, $A=0$, $\Big{(}x_0(0),\,\dot{x}_0(0)\Big{)}=(2,\,0)$
  (Appendix~\ref{app:sim}). The two approximate expressions match the simulated
  trajectory in panel~(a) over a time interval of one period
  $t \lesssim 2 \pi/\sqrt{b} \approx 6.28$. However, the asymmetric features of the
  oscillator's limit cycle are reproduced only by Eq.~(\ref{eq:xi1}).
}
\end{figure*}

The two-timing solution $\xi_1(t)$ reproduces better the asymmetric trajectory of
the Van der Pol limit cycle (Fig.~\ref{fig:one}) than $\xi_0(t)$. Both Eqs.~\eqref{eq:xi0}
and \eqref{eq:xi1} have the same frequency of oscillation $\omega=\sqrt{b}[1+ O(\mu^2)]$,
whose corrections are of quadratic order in the parameter $\mu$ \cite[Sec. 7.6]{StrogatzII}.
As discussed in the following Appendix, Eq.~\eqref{eq:xi0} is more convenient to
describe time-invariant statistics of the response terms $\xi(t)$ and $\gamma_\xi$
in Eq.~\eqref{eq:solution}, whereas Eq.~\eqref{eq:xi1} yields a more accurate expression
for the time autocorrelation function.

\section{\label{app:sta} Statistical properties of noisy relaxation oscillations}
Even in the absence of the external force $f(t)$ in Eq.~(\ref{eq:main}), relaxation
oscillations of the Van der Pol oscillator have nontrivial statistics. In the case
of the zeroth-order approximate solution $x_0(t) \approx \xi_0(t)$ [Eq.~(\ref{eq:xi0})]
we can find an exact probability distribution $p(\xi_0)$, which is given by the arcsine
law \cite[Chapters 16 and 17]{prbdrbs}---a special case of beta distributions with the
support interval shifted by $-1/2$ and scaled by $2\alpha$:
\begin{multline}
\label{eq:pxi}
  p(\xi_0) = \frac{d}{\pi d\xi_0} \left[
     \arcsin\left(\frac{\xi_0}{\alpha}\right) - \frac{\pi}{2}
  \right]
    = \frac{(\pi\alpha)^{-1}}{\sqrt{1 - \xi_0^2/\alpha}}.
\end{multline}
Statistics of $\xi_0(t)$ can also be evaluated by time averaging
\begin{align}\label{eq:varxi}
  \avg{\xi_0} = \int_0^{2\pi/\sqrt{b}} \frac{\sqrt{b}ds}{2\pi} \xi_0(s) = 0,
  \quad
  \avg{\xi_0^2} = \frac{\alpha^2}{2},
  \\\label{eq:acxi0}
  \avg{\xi_0(0) \xi_0(t)}
    = \int_0^{2\pi/\sqrt{b}} \frac{\sqrt{b}ds}{2\pi} \xi_0(s)\xi_0(s+t)
    \nonumber\\
    \qquad= \avg{\xi_0^2}\cos(\sqrt{b}t)&.
\end{align}

The probability density of $\xi_0(t)$ has two singularities at the ends of its support
interval $\xi_0=\pm\alpha$. Histograms of the time series $\xi_0(t)$, as well as
of $x_0(t)$, have two distribution modes near these points. In the companion paper
\cite{Belousov2019I} we have succeeded in fitting a bimodal probability density of
the noisy Duffing oscillator to an approximate expression that was derived from a
power series for the exponential family of random variables. This approach unfortunately
fails in the case of the noisy Van der Pol oscillator: such an expansion may not
exist near the two singularities at which $p(\xi_0)$ tends to infinity.

For the probability density of $x_\xi(t)$ [Eq.~(\ref{eq:solution})], regarded as
the sum of two independent variables $\xi(t)\simeq\xi_0(t)$ and $\tilde\gamma(t)$,
there is no simple analytical expression. However the Fourier image $\eta(X_\xi)$---the
characteristic function of $x_\xi$ for the reciprocal variable $X_\xi$---can be obtained
in a closed form. Because $\tilde\gamma(t)$ is Gaussian, we have
\begin{equation}\label{eq:chf}
  \eta(X_\xi) = \avg{\me^{\mi X_\xi x_\xi}}
    = J_0(\alpha X_\xi)\exp\left(-\frac{A^2 X_\xi^2}{4 \avg{a_\xi} \avg{b_\xi}}\right),
\end{equation}
in which
$$
  J_0(\alpha X_\xi) = \int_0^{2\pi/\sqrt{b}} \frac{\sqrt{b}ds}{2\pi} \me^{\mi X_\xi \xi_0(s)}
$$
is the characteristic function of $\xi_0(t)$ with $J_0(\cdot)$ being the zeroth-order
Bessel function of the first kind. Using $\xi(t)\simeq\xi_0(t)$ in Eqs.~\eqref{eq:avga}
and \eqref{eq:avgb} we get $\avg{a_\xi} \approx -a$ and $\avg{b_\xi} = b$.

\begin{figure*}[t]
\includegraphics[width=1\textwidth]{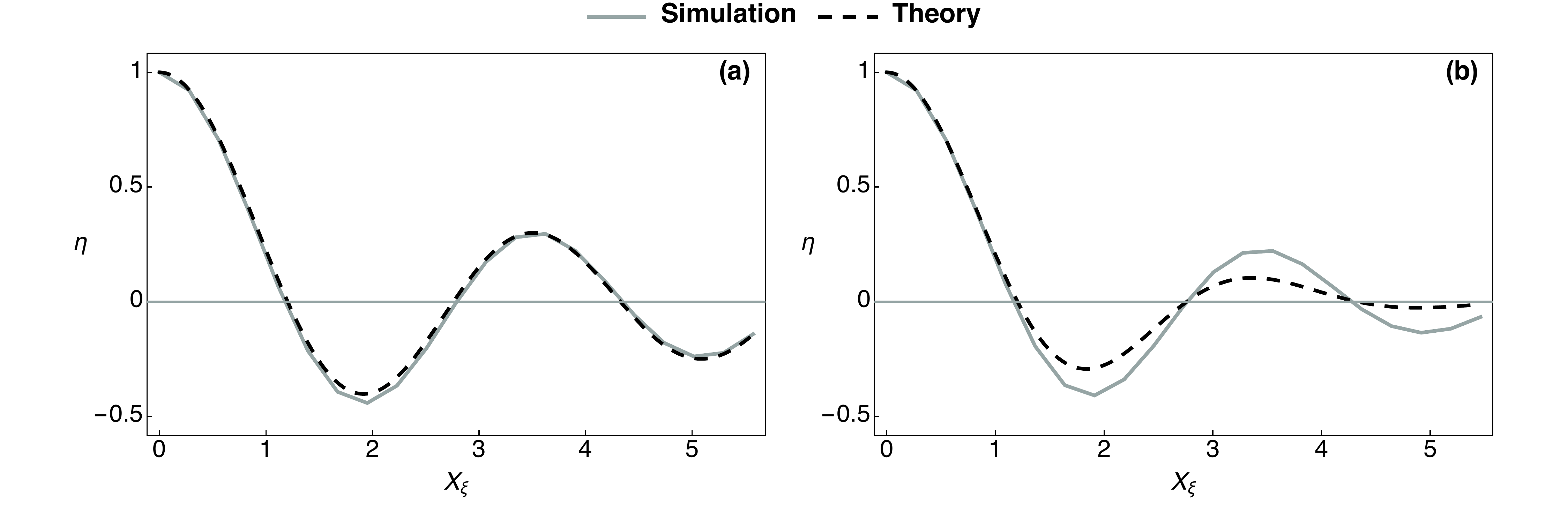}
\caption{\label{fig:chf}
  Comparison of the theoretical expression given by Eq.~(\ref{eq:chf}) with the empirical
  characteristic function $\eta(X_\xi)$ obtained by simulating the Van der Pol Eq.~\eqref{eq:main}.
  The system parameters are: (a) $\mu=1$, $A=0.0$ and (b) $\mu=1$, $A=0.6$. In these
  examples the theory is accurate at least for $X_\xi\lesssim\avg{x^2}^{-1/2} \approx 0.7$.
}
\end{figure*}

In Fig.~\ref{fig:chf} we compare the empirical characteristic function of $x(t)$
with Eq.~(\ref{eq:chf}) for two representative examples. Our analytical expression
for $\eta(X_\xi)$ is accurate at least for $X_\xi\lesssim\avg{x}^{-1/2}$ even for
large noise amplitudes $A$. The theory is in excellent agreement with the simulations
for $A\equiv0$.

Although we have not obtained analytical expressions analogous to Eqs.~\eqref{eq:pxi}
and \eqref{eq:chf} for the first-order approximate solution $\xi_1(t)$, its
autocovariance function can be evaluated by time averaging:
\begin{multline}
  \avg{\xi_1(0) \xi_1(t)}
    = \int_0^{2\pi/\sqrt{b}} \frac{\sqrt{b}ds}{2\pi} \xi_1(s)\xi_1(s+t)\\
    = \avg{\xi_1^2} \frac{\cos(\sqrt{b} t)}{1 + 5\mu^2/32}\left\{
      1 + \frac{\mu^2}{32}[4 + \cos(2\sqrt{b} t)]
    \right\},
\end{multline}
in which $\avg{\xi_1^2} = (1 + 5 \mu^2/32) \alpha^2/2$.

Our simulations show that the theoretical expressions based on the linear-order approximation
$\xi(t)\simeq\xi_1(t)$ overestimate the variance of the autonomous term $x_0(t)$,
as well as of the noisy oscillations $x(t)$. This discrepancy might be caused by a
broader orbit of $\xi_1(t)$ in the phase space, as compared to $x_0(t)$ and $\xi_0(t)$
[Fig.~\ref{fig:one}(b)]. Because the solution $\xi_0(t)$ provides a more accurate
estimate of the variance $\avg{x_0^2}$, Eq.~\eqref{eq:alphacA} is based on \eqref{eq:xi0}.

Because the terms $\xi(t)$ and $\tilde\gamma(t)$ in Eq.~\eqref{eq:solution} are statistically
independent, the autocorrelation function of $x_\xi(t)$ is given simply by
\begin{equation}\label{eq:chixi}
  \chi_\xi(t) = \frac{\avg{x_\xi(0) x_\xi(t)}}{\avg{x_\xi^2}}
    = \frac{
      \avg{\xi(0)\xi(t)} + \avg{\tilde\gamma(0) \tilde\gamma(t)}
    }{
      \avg{\xi^2} + \avg{\tilde\gamma^2}
    }.
\end{equation}
Approximating $\xi(t)$ by $\xi_0(t)$ and $\xi_1(t)$ we obtain, respectively,
\begin{eqnarray}\label{eq:chi0}
  \chi_0(t) &=& \frac{\avg{\xi_0^2}}{\avg{\xi_0^2} + \avg{\tilde\gamma_0^2}} \cos(\sqrt{b} t)
    + \frac{\avg{\tilde\gamma_0^2}}{\avg{\xi_0^2} + \avg{\tilde\gamma_0^2}}
    \nonumber\\
    &\times& \me^{-\frac{a_0 t}{2}}\left[
      \cos(\Omega_0 t) + \frac{a_0}{2\Omega_0} \sin(\Omega_0 t)
    \right],
  \\\label{eq:chi1}
  \chi_1(t) &=& \frac{\avg{\xi_1^2}}{\avg{\xi_1^2} + \avg{\tilde\gamma_1^2}}
    \times \frac{\cos(\sqrt{b} t)}{1 + 5\mu^2/32}
    \nonumber\\&\times& \left\{
      1 + \frac{\mu^2}{32}[4 + \cos(2\sqrt{b} t)]
    \right\}
    + \frac{\avg{\tilde\gamma_1^2}}{\avg{\xi_1^2} + \avg{\tilde\gamma_1^2}}
    \nonumber\\
    &\times& \me^{-\frac{a_1 t}{2}}\left[
      \cos(\Omega_1 t) + \frac{a_1}{2\Omega_1} \sin(\Omega_1 t)
    \right],
\end{eqnarray}
in which
\begin{eqnarray}
  a_0 = \mu \sqrt{b}, &\quad& a_1 = \mu \sqrt{b} (1+5\mu^2/16),\\
  \Omega_0 = \sqrt{b^2-a_0^2/4},
    &\quad& \Omega_1 = \sqrt{b^2-a_1^2/4},\\
  \avg{\tilde\gamma_0^2} = \frac{A^2}{2 a_0 b}
    &\quad& \avg{\tilde\gamma_1^2} = \frac{A^2}{2 a_1 b}.
\end{eqnarray}

\begin{figure}[t]
\includegraphics[width=1\columnwidth]{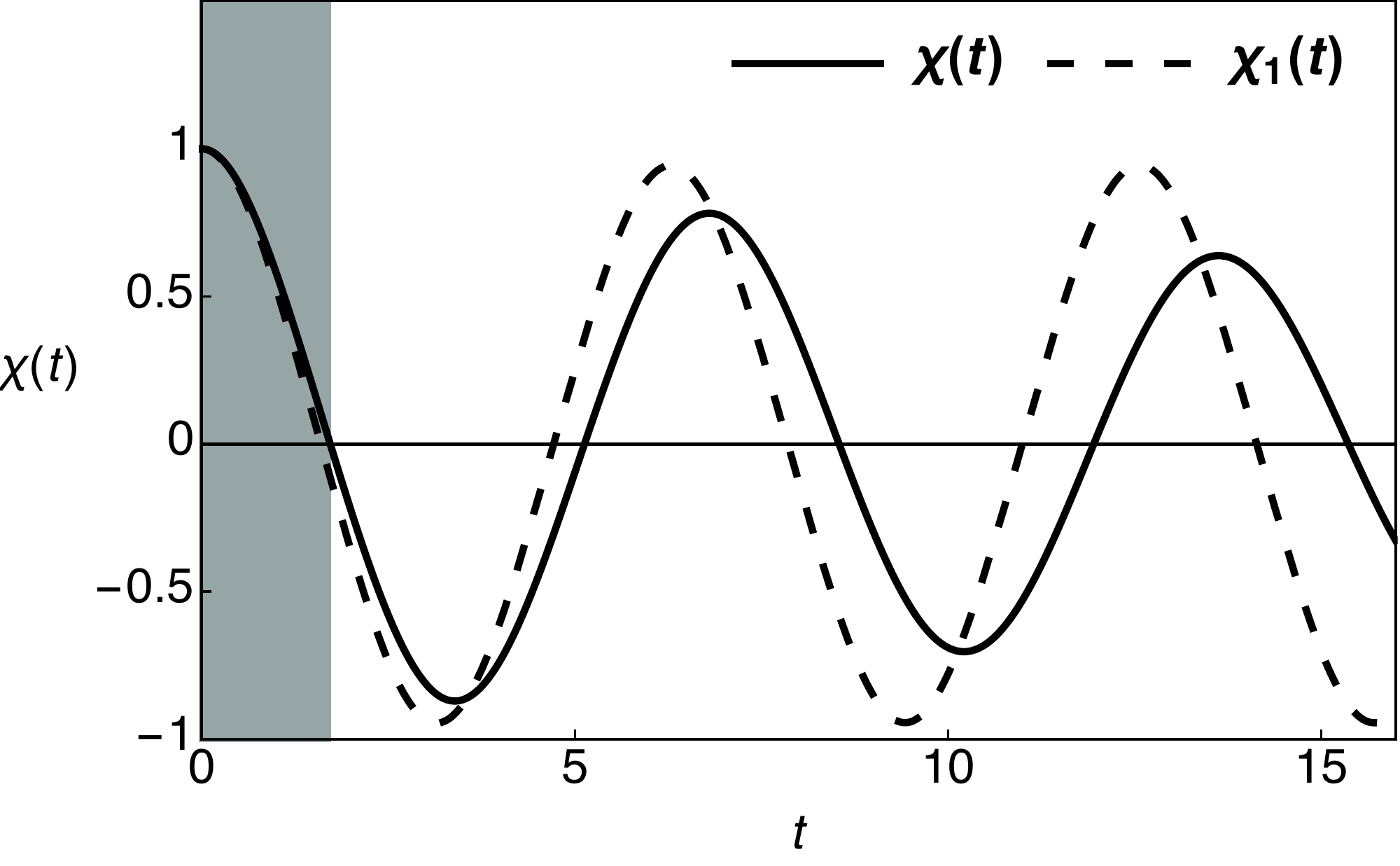}
\caption{\label{fig:chi}
  Comparison of the theoretical expression $\chi_1(t)$ [Eq.~\eqref{eq:main}] with
  the empirical time autocorrelation function $\chi(t)$ obtained from simulations
  of the Van der Pol Eq.~\eqref{eq:main}. The system parameters are $\mu=1$, $A=0.6$.
  The agreement is good in the shaded region selected by the criterion of Lagarkov
  and Sergeev (Sec.~\ref{sub:pinf}).
}
\end{figure}

In Fig.~\ref{fig:chi} the empirical autocorrelation function $\chi(t)$ obtained from
simulations of Eq.~\eqref{eq:main} is compared with the theoretical curve $\chi_1(t)$.
To avoid redundancy, we do not reproduce the plot of $\chi_0(t)$ [Eq.~\eqref{eq:chi0}],
which is almost indistinguishable from that of $\chi_1(t)$. Our theoretical expression
predicts well the initial decay of the empirical time autocorrelation function, although
a moderate phase difference accumulates at longer times.

As discussed in Sec.~\ref{sub:lrsp}, the approximate Eq.~\eqref{eq:solution} does
not account for the fluctuating phase shift of the noisy oscillations that occur
in presence of the driving force $f(t)$ and vanish as $A\to0$. These stochastic phase
variations accumulate slowly and decorrelate the time series of $x(t)$. Consequently
the empirical autocorrelation function $\chi(t)$ decays to zero as $t\to\infty$
(Fig.~\ref{fig:chi}) unless $f(t)\equiv0$. This decay becomes faster as the noise
amplitude $A$ increases. The persistent periodic terms $\propto\cos^2(\sqrt{b}t)$
in Eqs.~\eqref{eq:chi0} and \eqref{eq:chi1}, whose amplitude is constant, are therefore
accurate only at small time scales.

Although graphs of Eqs.~\eqref{eq:chi0} (not reproduced in Fig.~\ref{fig:chi}) and
\eqref{eq:chi1} are indistinguishable, fitting the latter expression to empirical
time autocorrelations (Sec.~\ref{sub:pinf}) performs better, because it provides
tighter constraints on the parameter $\mu$. Fitting Eq.~\eqref{eq:chi0}, in which
the first term  $\propto \cos^2(\sqrt{b} t)$ depends only on the parameter $b$, yields
less accurate estimates of the constant $\mu$.

\section{\label{app:sim}Simulation algorithm}
In our computational experiments we use a companion system of Eq.~(\ref{eq:main})
with $\bm{X} = (x,y) = (x,\dot{x})$:
\begin{equation}
  \begin{cases}
    \dot{x} = y\\
    \dot{y} =  -(a + c x^2) y - b x + f(t)
  \end{cases}.
\end{equation}
We adopt a second-order operator-splitting approach \cite{Tuckerman1992} for stochastic
systems \cite[Appendix~C]{Belousov2019I,Belousov2017}, by decomposing the time-evolution
operator $\oper{T}$ as
\begin{equation}\label{eq:split}
  \dot{\bm{X}} = \oper{T} \bm{X} = (
    \oper{T}_f + \oper{T}_{y} + \oper{T}_{x}
  ) \bm{X},
\end{equation}
in which
\begin{align*}
  \oper{T}_x = y \bm{\partial_x},\;
  \oper{T}_y = -(a + c x^2) y \bm{\partial_y},\;
  \oper{T}_f = (f - b x) \bm{\partial_y}.
\end{align*}

The formal solution of Eq.~(\ref{eq:split}) for a time step $\Delta{t}$
$$ \bm{X}(t+\Delta{t}) = \exp(\oper{T} \Delta{t}) \bm{X}(t) $$
can be approximated by
\begin{multline}\label{eq:operator}
  \exp[\oper{T}\Delta{t} + \mathcal{O}(\Delta{t}^2)] =
      \exp\left(\frac{\oper{T}_x\Delta{t}}{2}\right)
      \exp\left(\frac{\oper{T}_y\Delta{t}}{2}\right)\\
      \times\exp(\oper{T}_f \Delta{t})
      \exp\left(\frac{\oper{T}_y\Delta{t}}{2}\right)
      \exp\left(\frac{\oper{T}_x\Delta{t}}{2}\right)
    .
\end{multline}
The action of individual operators of the form $\exp(\oper{L}\Delta{t})$ is determined
by the differential equation
\begin{equation}
  \dot{\bm{X}}(t) = \oper{L} \bm{X}(t) \Rightarrow \bm{X}(t+\Delta{t})
    = \exp(\oper{L}\Delta{t}) \bm{X}(t).
\end{equation}
The operators $T_x$, $T_y$, and $T_f$ produce linear equations of the above type
and their action is given by
\begin{align*}
  &\me^\frac{t \oper{T}_x}{2} (x,y) = \left(x + y\frac{\Delta{t}}{2}, y\right),\\
  &\me^\frac{t \oper{T}_y}{2} (x,y) =
    \left(x, y \me^{-\frac{(a + c x^)\Delta{t}}{2} }\right),\\
  &\me^{t \oper{T}_f} (x,z)
    = \left(x, y + b x^2 \Delta{t} + \int_0^{\Delta{t}} ds\, f(s)\right).
\end{align*}
\newpage
For each value of the parameter $A$, the simulations reported in Sec.~\ref{sec:num}
involved $5\times10^5$ time steps of a size $\Delta{t}=0.01$ in the reduced units.
Statistics were calculated from single trajectories sampled at time intervals of
$0.05$, which included $10^5$ observations.

\section{\label{app:hcb} Hidden Van der Pol model for hair-bundle oscillations}

In this appendix we reduce the parsimonious model of hair-bundle oscillations
\cite{ONH2012} to a simple form of a linear oscillator $x(t)$ coupled to a hidden
Van der Pol oscillator $z(t)$ [Eqs.~\eqref{eq:xz} and \eqref{eq:ddz}]. The starting
point of our derivation is a system of equations for $(x,y)$ \cite[\textit{cf.} Eqs.~(1) and (S6)]{ONH2012}:
\begin{eqnarray}\label{eq:ddx}
  M \ddot{x} &=& -\Gamma \dot{x} - K_0 x + K_1 (x-y)
    \nonumber\\&-& B_0 (x-y)^3 + F,
  \\\label{eq:dy}
  \tau_0 \dot{y} &=& K_2 x - K_3 y,
\end{eqnarray}
in which $M$ and $x(t)$ are the mass and the position of the hair bundle, $y(t)$
is an adaptation coordinate, $K_0>0$, $K_1 > 0$, $K_2 > 0$, $K_3>0$, and $B_0>0$
are elastic constants, whereas $\Gamma > 0$, $\tau_0 > 0$, and $F(t)$ are, respectively,
a friction coefficient, a relaxation time of the adaptation coordinate $y$, and an
external force.

We proceed by taking the overdamped limit of Eq.~\eqref{eq:ddx} $M/\Gamma \to 0$
and substituting a simplifying assumption $K_2 \approx K_3$ into \eqref{eq:dy}:
\begin{eqnarray}\label{eq:simplex}
  \dot{x} &=& - k_0 x + k_1 (x-y) - b_0 (x-y)^3 + f,
  \\\label{eq:simpley}
  \dot{y} &=& c_x (x - y),
\end{eqnarray}v,
in which $k_{i=0,2} = K_{i=0,1}/\Gamma$, $c_x = K_2/\Gamma$, $b_0 = B_0/\Gamma$,
and $f(t) = F(t)/\Gamma$. Next we use a substitution of variables $z = x-y$, which
transforms Eqs.~\eqref{eq:simplex} and \eqref{eq:simpley} into
\begin{eqnarray}
  &&\dot{x} = - k_0 x + k_1 z - b_0 z^3 + f,
  \\\label{eq:xdzz}
  &&\dot{x} - \dot{z} = c_x z.
\end{eqnarray}
By subtracting the second of the above equations from the first one, we can recast
our original system $(x,y)$ into a system $(x,z)$:
\begin{eqnarray}\label{eq:dx}
  \dot{x} &=& - k_0 x + k_1 z - b_0 z^3 + f,
  \\\label{eq:dz}
  \dot{z} &=& - k_0 x - (c_x - k_1) z - b_0 z^3 + f.
\end{eqnarray}
Further, we use Eq.~\eqref{eq:dz} and its time derivative to express $k_0 x$ and
$k_0 \dot{x}$ as
\begin{eqnarray}
  k_0 x &=& -\dot{z} - (c_x - k_1) z - b_0 z^3 + f,\\
  k_0 \dot{x} &=& -\ddot{z} - (c_x - k_1) \dot{z} - 3 b_0 z^2 \dot{z} + \dot{f}.
\end{eqnarray}
Then we use the above equations to eliminate $\dot{x}$ and $x$ from Eq.~\eqref{eq:dx}
and thus obtain
\begin{equation}\label{eq:vdp}
  \ddot{z} + (k_0 - k_1 + c_x) \dot{z} + k_0 c_x z + 3 b_0 z^2 \dot{z} = \dot{f}.
\end{equation}
If in the last equation we identify
$$
  a = k_0 - k_1 + c_x,\quad b = k_0 c_x,\quad c = 3 b_0,\quad f=a\dot{w},
$$
we obtain Eq.~\eqref{eq:ddz}---the hidden Van der Pol oscillator. Equation~\eqref{eq:xdzz}
entails the linear coupling between $x(t)$ and $z(t)$ [Eq.~\eqref{eq:xz}].

In simulations we integrate Eqs.~\eqref{eq:dx} and \eqref{eq:dz} using a
decomposition of the time evolution operator $\oper{T}$:
\begin{equation}
  (\dot{x},\dot{z}) = \oper{T} (x,z)
    = (\oper{T}_f + \oper{N}_z + \oper{L}_z + \oper{L}_x) \bm{X},
\end{equation}
in which
\begin{align}
  \oper{L}_x =& (k_1 z - b_0 z^3 - k_0 x) \bm{\partial_x},\\
  \oper{L}_z =& -(k_0 x  + k_z z) \bm{\partial_z},\\
  \oper{N}_z =&  -b_0 z^3 \bm{\partial_z},\;
  \oper{T}_f = f (\bm{\partial_x} + \bm{\partial_z}),\;
\end{align}
and $k_z = c_x - k_1$. Up to the second order in $t$ we then have
\begin{equation}\label{eq:step1}
  \me^{t \oper{T} + \mathcal{O}(t^2)} =
    \me^\frac{t \oper{L}_x}{2}\me^\frac{t \oper{L}_z}{2}
    \me^\frac{t \oper{N}_z}{2} \me^{t \oper{T}_f} \me^\frac{t \oper{N}_z}{2}
    \me^\frac{t \oper{L}_z}{2}\me^\frac{t \oper{L}_x}{2},
\end{equation}
in which the action of individual operators is
\begin{align*}
  \me^\frac{t \oper{L}_x}{2} (x,z) =& \left(
      x \me^{-\frac{k_0 t}{2}} + \frac{k_1 z - b_0 z^3}{k_0}\left(1-\me^{-\frac{k_0 t}{2}}\right),
    z \right),\\
  \me^\frac{t \oper{L}_z}{2} (x,z) =& \left( x,
      z \me^{-\frac{k_z t}{2}} - \frac{k_0 x}{k_z}\left(1-\me^{-\frac{k_z t}{2}}\right)
  \right),\\
  \me^\frac{t \oper{N}_z}{2} (x,z)
    =& \left(x, \frac{z}{\sqrt{1 + z^2 b_0 t}}\right),
  \\\label{eq:eTf1}
  \me^{t \oper{T}_f} (x,z)
    =& \left(
      x+\int_0^t ds\, f,
      z+\int_0^t ds\, f
    \right).
\end{align*}

As proposed in Sec.~\ref{sec:exp}, the dynamical Eqs.~\eqref{eq:simplex}--\eqref{eq:vdp}
can be imitated by a driven harmonic oscillator. To implement this idea we replace
the cubic nonlinear term of the original problem by a deterministic active force
$f_a(t) = \alpha_a \cos(\omega_a t) + \beta_a \sin(\omega_a t)$ and a compensatory
linear term $\Delta{k}(x-y)$ with real constants $\alpha_a$, $\beta_a$, $\omega_a$,
$\Delta{k}$:
\begin{eqnarray}\label{eq:hx}
  \dot{x} &=& - k_0 x + k_{xy} (x-y) + f_a + f,
  \\\label{eq:hy}
  \dot{y} &=& c_x (x - y),
\end{eqnarray}
in which $k_{xy} = k_1 + \Delta{k}$. By requiring that the steady-state solution of
the above system is given by Eq.~\eqref{eq:zeta}, we find
$$
  \alpha_a = \mu \sqrt{b}\alpha,\;
  \beta_a = 0,\;
  \omega_a = \sqrt{b},\;
   k_{xy} = k_0 + c_x - \mu\sqrt{b}.
$$
Instead of Eq.~\eqref{eq:vdp} we then obtain
\begin{equation}\label{eq:zho}
  \ddot{z} + \mu\sqrt{b} \dot{z} + b z = \dot{f}_a + \dot{f},
\end{equation}
whereas the linear coupling Eq.~\eqref{eq:xz} remains unaltered.

\bibliographystyle{apsrev4-1}
\bibliography{refs}

\begin{thebibliography}{47}
\bibitem [{foo()}]{foot1}%
  \BibitemOpen
  \href@noop {} {\bibinfo  {journal} {These experiments were conducted in a
  sample chamber that was cooled down to 285 K by a peltier element and in
  accordance with the policies of The Rockefeller University's Institutional
  Animal Care and Use Committee (IACUC Protocol 16942)}\ }\BibitemShut
  {NoStop}%
\end{thebibliography}
\end{document}